\documentclass[conference]{IEEEtran}
\IEEEoverridecommandlockouts

\usepackage{cite}
\usepackage{amsmath,amssymb,amsfonts}
\usepackage{algorithmic}

\usepackage{textcomp}
\usepackage{xcolor}
\usepackage{graphicx}
\usepackage{booktabs}
\usepackage{makecell}
\usepackage{multirow}
\usepackage[T1]{fontenc}
\usepackage[utf8]{inputenc}
\usepackage{url}
\def\BibTeX{{\rm B\kern-.05em{\sc i\kern-.025em b}\kern-.08em
    T\kern-.1667em\lower.7ex\hbox{E}\kern-.125emX}}
\renewcommand{\footnoterule}{%
  \kern -3pt
  \hrule width 0.4\columnwidth
  \kern 2.6pt
}
\begin{document}

\title{JASPER: Joint Audio and Speech Pre-trained Encoder Representations

}

\author{
\IEEEauthorblockN{
Geeth George\textsuperscript{*},
Ameenudeen P E\textsuperscript{*},
Hrishikesh H Pillai,
Sriram Ganapathy \\
}
\IEEEauthorblockA{
LEAP Laboratory, Electrical Engineering,
Indian Institute of Science, Bangalore, India\\
\texttt{geethgeorge@iisc.ac.in, ameenudeenp@iisc.ac.in}
}
}

\maketitle
\begingroup
\renewcommand{\thefootnote}{}
\footnotetext{\textsuperscript{*}Equal contribution.}
\endgroup

\begin{abstract}
Self-supervised learning (SSL) for speech and audio has largely progressed along separate tracks: speech models emphasize time-domain prediction, whereas audio representation learning has focused  on time--frequency patterns. This separation creates a compatibility gap, limiting cross-domain generalization. In this work, we introduce \textbf{JASPER},   \textbf{J}oint \textbf{A}udio and \textbf{S}peech \textbf{P}re-trained  \textbf{E}ncoder    \textbf{R}epresentations, a framework that augments speech-pretrained models with time--frequency objectives.
Specifically, JASPER performs masked prediction of temporal and spectral targets over long audio segments, enabling spectro-temporal representation learning of speech and audio signals. The proposed method consistently outperforms multiple baselines and existing speech/audio encoders on  diverse speech, audio, and music tasks, demonstrating the effectiveness of unified spectro-temporal modeling.
\end{abstract}


\section{Introduction}
Self-supervised learning (SSL) has emerged as a dominant paradigm for learning robust representations from unlabeled acoustic data. Existing SSL-based audio models can broadly be categorized into two families based on their input modality: raw waveform-based models and spectrogram-based models. Waveform-based models operate directly on time-domain audio signals, learning hierarchical representations without relying on handcrafted acoustic features. Examples include contrastive predictive frameworks such as wav2vec~2.0 \cite{wav2vec2},  clustering-based masked prediction methods such as HuBERT \cite{hubert} and WavLM \cite{chen2022wavlm}. SSL methods based on the efficient masked autoencoder framework such as Dasheng \cite{dinkel2024scaling}, as well as large-scale weakly-supervised models such as Whisper \cite{radford2023robust} also provide alternative design choices for audio encoders. For example, the Qwen-audio model uses the Whisper encoder as the audio input processing framework which is fine-tuned for speech and audio tasks \cite{chu2024qwen2}.
While such waveform-based encoders achieve state-of-the-art performance on speech-centric tasks such as ASR, speaker verification, and speech translation, they fail to generalize to diverse non-speech audio tasks \cite{ultras}.

Spectrogram-based models, in contrast, operate on time--frequency representations such as log-Mel spectrograms. Models such as the Audio Spectrogram Transformer (AST) \cite{ast} and Self-Supervised Audio Spectrogram Transformer (SSAST) \cite{ssast} adopt patch-based tokenization and global self-attention mechanisms to model long-range dependencies across time and frequency. Spectrogram-based encoders demonstrate strong performance on general audio tasks \cite{chen2022beats}.

Despite their strong performance within their intended domains, the distinct design choices of speech and audio models limit their cross-domain applicability. Frame-level tokenizers used in speech models are less effective at representing complex environmental sounds, whereas patch-based audio models often fall short on speech tasks that depend on precise temporal and sequential information.

While prior works \cite{ultras} have shown that incorporating temporal learning into spectrogram transformers improves speech performance, the complementary direction remains underexplored. To address these challenges, we propose JASPER - \textbf{J}oint \textbf{A}udio and \textbf{S}peech \textbf{P}re-trained \textbf{E}ncoder \textbf{R}epresentations. JASPER treats speech and audio pre-training as complementary spectro-temporal objectives over a shared Transformer encoder, jointly predicting temporal targets and spectral patch targets over long audio contexts. 

\begin{figure*}[t!]
    \centering
    \includegraphics[width=\linewidth]{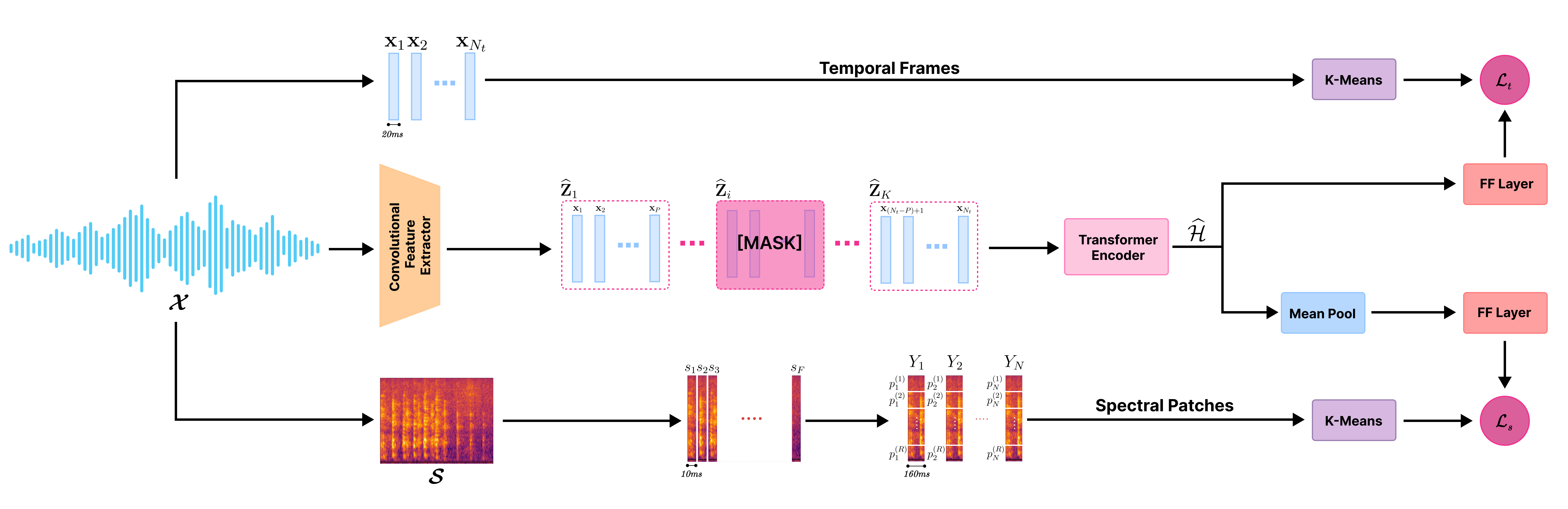}
    \vspace{-1em} 
    \caption{Block schematic of the proposed JASPER framework for joint spectro-temporal modeling of audio data. ($N_t$ = 400, F = 800, $N$ = 50, $R$ = 8).}
    \label{fig:block-schematic}
    \vspace{-0.1in}

\end{figure*}
To validate the embedding quality of JASPER and identify the factors contributing to  performance gains, we introduce a comprehensive, multi-tiered evaluation framework. First, we apply the frozen representations to traditional intrinsic benchmarking using SUPERB-style \cite{superb} downstream linear evaluations, ensuring a direct assessment of  acoustic and phonetic properties. Second, we evaluate the extrinsic utility of the embeddings using an encoder-centric evaluation with the X-ARES benchmark \cite{zhang2025x} and also by deploying JASPER as the unified front-end acoustic encoder for a Large Audio Language Model via the XARES-LLM\cite{interspeech2026_aecc}  framework.
In these experiments, we highlight the superior performance and generalization of JASPER compared to various other baseline representations. 

The main contributions of this work are summarized below:
\begin{itemize}
    \item We propose JASPER, a unified self-supervised audio encoder that combines waveform-based temporal learning with spectrogram-based spectral learning to improve generalization across speech, environmental audio, and music tasks.
    \item We introduce a {long-context masking strategy} over waveform-derived latent representations, where contiguous temporal segments are masked to encourage the encoder to infer missing acoustic content from broader temporal context.

    \item We propose a {spectral prediction objective} for waveform encoders by constructing time-aligned spectral patch targets from log-Mel representations. The model is trained to predict both temporal pseudo-labels and frequency-localized spectral targets, thereby injecting time-frequency supervision into a raw-waveform encoder without requiring spectrogram input at inference time.

    \item We establish JASPER as a universal audio encoder through broad speech, audio, music, and audio-LLM evaluations, demonstrating robust representations for both discriminative and generative audio understanding.
\end{itemize}

\section{JASPER Encoder Framework}
Figure \ref{fig:block-schematic} illustrates the workflow of JASPER consisting of  a self-supervised audio encoder designed to learn unified representations for both speech and general audio signals. 
\subsection{Input Pre-Processing}

Let $\boldsymbol{\mathcal{X}} \in \mathbb{R}^{16000 \cdot T \times 1}$ denote a raw speech/audio recording of duration $T$ seconds, sampled at $16$ kHz. 
We process $\boldsymbol{\mathcal{X}}$ using a convolutional feature encoder which has an overall stride of $320$, resulting in a sequence of latent temporal representations at $50$Hz:

\begin{equation}
\mathbf{X} = [\mathbf{x}_1, \mathbf{x}_2, \dots, \mathbf{x}_{N_t}], 
\quad \mathbf{x}_i \in \mathbb{R}^{D_t},
\end{equation}
where $N_t = 50T$ denotes the total number of temporal frames and $D_t$ is the dimension of the convolutional  encoder.

\subsection{Temporal Segmentation}
The raw speech utterance $\boldsymbol{\mathcal{X}}$ 
is divided into non-overlapping temporal segments, each of duration $\tau$ seconds. 
This segmentation induces a corresponding division of the latent feature sequence 
$\mathbf{X}$ into non-overlapping windows of $P$ frames, where
$P = 50\tau$. 
Let the windowed latent representation be denoted as
\[
\boldsymbol{\mathcal{Z}} = [\mathbf{Z}_1, \mathbf{Z}_2, \dots, \mathbf{Z}_K].
\]
Here, $\mathbf{Z}_k \in \mathbb{R}^{P \times D_t}$ represents the $k$-th windowed latent segment, and 
$K = \frac{T}{\tau}$.  $\mathbf{Z}_k$ is constructed by grouping $P$ consecutive latent vectors, i.e.,
\[
\mathbf{Z}_k =
\left[
\mathbf{x}_{(k-1)P + j}
\right]_{j=1}^{P},
\quad k = 1, \dots, K.
\]
\subsection{Long-Context masking}
To enable self-supervised learning, we apply a   masked modeling strategy 
to the windowed latent representations $\boldsymbol{\mathcal{Z}}$. 
Each latent window is independently masked with probability $p$. 
Furthermore, to encourage the model to capture longer temporal dependencies, 
if a window is selected for masking, its subsequent window is also masked 
with a fixed probability $p'$.
\\
Let the resulting masked latent sequence be denoted as:
\[
\boldsymbol{\widehat{\mathcal{Z}}} 
= [\widehat{\mathbf{Z}}_1, \widehat{\mathbf{Z}}_2, \dots, \widehat{\mathbf{Z}}_K].
\]
The masking operation is defined as:
\[
\widehat{\mathbf{Z}}_k =
\begin{cases}
\mathbf{[MASK]}, & \text{if location } k \text{ is selected for masking}, \\
\mathbf{Z}_k, & \text{otherwise},
\end{cases}
\]
where $\mathbf{[MASK]} \in \mathbb{R}^{P \times D_t}$ denotes a learnable mask embedding 
(or a zero mask of the same dimensionality).
In our experiments, we use $p = 0.6$ and $p' = 0.2$.

\subsection{Transformer Encoding of Temporal Frames}
To retain temporal order information, a convolutional positional embedding 
is added to the frame sequence within each segment. The resulting sequence 
is then passed to a transformer encoder, which produces contextualized 
temporal representations, each of dimension $D'$:
\[
\mathbf{\widehat{H}}_k =
\left[
\mathbf{\widehat{h}}_k^{(1)}, 
\mathbf{\widehat{h}}_k^{(2)}, 
\dots, 
\mathbf{\widehat{h}}_k^{(P)}
\right],
\quad k = 1, \dots, K \quad \mathbf{\widehat{h}}_k^{(i)} \in \mathbb{R}^{D'}. 
\]
\subsection{Temporal Targets and Loss Function}

For each $\mathbf{x}_i$, 
a discrete pseudo-label $\mathbf{C}_{t,k}^{(i)}$ is obtained by clustering the frame-level features $\mathbf {x}_i$ using K-means clustering. 
We define a temporal loss function $\mathcal{L}_{\texttt{t}}$  that encourages the transformer encoder of the JASPER model to predict these pseudo-labels. The loss is computed  only on the masked frames.

Specifically, a K-means quantizer 
$\mathcal{Q}$ is applied to frame-level embeddings to generate 
cluster assignments:
\[
\textbf{C}_{t,k}^{(i)} = \mathcal{Q}(\mathbf{x}_i), \quad \textbf{C}_{t,k}^{(i)}\in \{1, 2, \dots, V_t\} 
\]
\[
 k = 1,\dots ,K \quad  \quad i = 1, \dots ,P \]
where $\textbf{C}_{t,k}^{(i)}$ denotes the cluster label corresponding to the $i$-th frame in the $k$-th window and $V_t$ denotes the number of clusters.
Given the masked  sequence ($\boldsymbol{\widehat{\mathcal{Z}}}$) as input, the transformer encoder 
produces contextualized representations for all temporal frames. 
Let ${\mathbf{\widehat{h}}_{k}^{(i)}} \in \mathbb{R}^{D'}$ denote the output 
corresponding to the $i$-th frame of the $k$-th window.
Each frame representation is projected using a learnable feed-forward layer,  with parameters $\mathbf{W}_t \in \mathbb{R}^{V_t \times D'}$, followed by a softmax operation:
\[
\mathbf{\widehat{\textbf{p}}}_k^{(i)} = \text{softmax}(\mathbf{W}_t \mathbf{\widehat{h}}_{k}^{(i)}).
\]
The training objective 
is computed only over masked frame indices. The temporal loss is defined as:

\begin{equation}\label{eqn:spectral-loss}
\mathcal{L}_{\texttt{t}} = \frac{1}{|\mathcal{M}|} \sum_{k \in \mathcal{M}} \sum_{i=1}^P\text{C.E.}  \left( \widehat{\mathbf{p}}_{k}^{(i)}, \textbf{C}_{t,k}^{(i)} \right)
\end{equation}
where $\mathcal{M}$ denotes the set of masked window indices, and $\text{C.E.}(\cdot, \cdot)$ denotes the standard cross-entropy loss.

\subsection{Spectral Targets and Loss}

Let the log-Mel spectrogram of waveform $\boldsymbol{\mathcal{X}}$ be
\[
\boldsymbol{\mathcal{S}} = [\mathbf{s}_1, \dots, \mathbf{s}_F],
\]
where $\mathbf{s}_i \in \mathbb{R}^{D}$, $D=128$, extracted using $128$-D Mel-filterbank features ($25$ ms window, $10$ ms hop). The spectrogram is divided into non-overlapping segments of $Q$ consecutive spectral frames corresponding to the temporal duration $\tau$ seconds :
\[
\boldsymbol{\mathcal{Y}} = [\mathbf{Y}_1, \dots, \mathbf{Y}_N],
\quad 
\mathbf{Y}_n \in \mathbb{R}^{Q \times D},
\quad 
N = \frac{F}{Q}.
\]
Each spectrogram segment $\mathbf{Y}_n$ captures the spectral 
structure of the windowed signal of $\tau$ seconds duration. Each $\mathbf{Y}_n$ is further split vertically into $R$ non-overlapping $Q \times Q$ square patches:
\[
\mathbf{P}_n =
\begin{bmatrix}
\mathbf{p}_n^{(1)} &
\mathbf{p}_n^{(2)} &
\cdots &
\mathbf{p}_n^{(R)}
\end{bmatrix}^{\!\top},~~
\mathbf{p}_n^{(r)} \in \mathbb{R}^{Q \times Q},
~~
R = \frac{D}{Q}.
\]
Spectral targets are obtained by flattening each patch and quantizing via a K-means quantizer $Q_s$:
\begin{gather*}
C_{s,k}^{(r)} = Q_s\big(\mathrm{vec}(p_n^{(r)})\big) \\[5pt]
C_{s,k}^{(r)} \in \{1,\dots,V_s\}, n = 1,\dots,N, r = 1,\dots,R.
\end{gather*}
Here, $V_s$ denotes the number of spectral K-means clusters. Consider a temporal segment-level representation for the transformer embeddings via mean pooling $
\bar{h}_k = \frac{1}{P} \sum_{i=1}^{P} \widehat{h}_k^{(i)}.
$ 
We design $R$ independent feed-forward heads $\{W_s^{(r)}\}_{r=1}^{R}$ to predict spectral cluster labels as:
\[
p_{s,k}^{(r)} =
\mathrm{softmax}(W_s^{(r)} \bar{h}_k).
\]
The spectral loss is computed over masked windows:
\[
L_s =
\frac{1}{|M|R}
\sum_{k \in M}
\sum_{r=1}^{R}
\mathrm{C.E.}\big(
p_{s,k}^{(r)}, C_{s,k}^{(r)}
\big).
\]
where C.E. is the standard cross-entropy loss and $\lvert M \rvert$ is the number of masked frames.

\subsection{Combined Training Objective}
The overall training objective of the model  is the sum of spectral loss $\mathcal{L}_{\text{s}}$ and  temporal loss $\mathcal{L}_{\text{t}}$:
\begin{equation}
\mathcal{L}_{\text{total}} =  \lambda \mathcal{L}_{\texttt{t}}  + (1-\lambda) \mathcal{L}_{\texttt{s}}
\label{lambda_eqn}
\end{equation}

\section{Implementation Details}
\subsection{Model Architecture and Targets}

Each input audio signal is truncated or zero-padded to $8$ s and the audio segment duration $\tau=0.16$ s is chosen to match syllable-length units.
 The raw waveform is processed by a convolutional feature encoder with $D_t = 512$  and  passed through a $12$-layer transformer encoder with $12$ attention heads, and $D'=768$. For self-supervised target construction, we use $V_t=500$ and $V_s=500$. 
 
For the spectral processing, we have $F = 800$ and $Q = 16$. This gives $N = 50$ non-overlapping windows. The number of vertical patches in each spectral segment is $R = 8$.

\subsection{Pre-training Setup and Optimization}
JASPER is pretrained under two data regimes: 200-hour and 2000-hour settings, each comprising equal amounts of speech and audio from LibriSpeech \cite{librispeech} and AudioSet \cite{gemmeke2017audio}. The encoder is first initialized using temporal-only targets, followed by joint predictive pre-training with combined temporal and spectral objectives for $1$M steps. Training uses the AdamW optimizer with weight decay of $0.05$, a constant learning rate of $2\times 10^{-5}$, and coefficients $\beta = [0.9, 0.98]$.

\begin{table*}[t!]
\caption{Downstream performance comparison under 200-hour and 2000-hour pre-training setups. All models are trained using the same data and number of pre-training steps. Results are reported as classification accuracy (\%).
}
\label{tab:1}
\vspace{-1.8em}
\begin{center}
\small
\setlength{\tabcolsep}{4.2pt} % Adjusted for perfect 11-column fit
\begin{tabular}{|l||c|c|c|c||c|c|c|c|c|r|}
\midrule
\midrule
\multirow{2}{*}{\textbf{Model}} & \multicolumn{4}{c||}{\textbf{Audio Tasks}} & \multicolumn{5}{c||}{\textbf{Speech Tasks}} \\
\cmidrule{2-10}
& \textbf{ESC-50} & \textbf{US8k} & \textbf{GTZAN} & \textbf{FMA\_SMALL} & \textbf{IEMOCAP} & \textbf{VoxC1} & \textbf{SPCV2} & \textbf{CREMA-D} & \textbf{LibriC.} \\
\midrule
\midrule
\multicolumn{10}{|c|}{\textbf{200 hour setup}} \\
\midrule
SSAST\cite{ssast}+MLM  & 77.65 & 78.46 & 80.08 & \textbf{53.75} & 60.49 & 32.19 & 78.89 & 54.60 & \textbf{68.77}  \\

HuBERT\cite{hubert}     & 66.10 & 70.05 & 79.88 & 48.12 & 59.91 & 28.83 & 84.82 & 60.35 & 68.18  \\
JASPER     & \textbf{83.55} & \textbf{78.84} & \textbf{89.69} & 52.5 & \textbf{65.02} & \textbf{53.93} & \textbf{92.80} & \textbf{70.01} & 66.91 \\
\midrule
\midrule
\multicolumn{10}{|c|}{\textbf{2000 hour setup}} \\
\midrule
SSAST\cite{ssast}+MLM  & 85.50 & \textbf{84.27} & 91.29 & 49.11 & 62.05 & 47.95 & 82.03 & 70.49 & 70.73 \\
HuBERT\cite{hubert}    & 80.55 & 78.42 & 91.89 & 49.25 & 66.27 & 64.56 & \textbf{95.55} & 68.46 & 67.55 \\
JASPER  &\textbf{ 88.05} & 83.66 & \textbf{91.99} & \textbf{53.75} & \textbf{68.76} & \textbf{91.18} & 94.78 & \textbf{76.89} & \textbf{72.13}  \\
\midrule
\midrule
\multicolumn{10}{l}{\footnotesize \textbf{US8k}: UrbanSound8K;  \textbf{FMA}: Free Music Archive; \textbf{VoxC1}: VoxCeleb1; \textbf{SPCV2}: Speech Commands V2;
\textbf{LibriC.}: LibriCount} \\

\end{tabular}
\end{center}
\label{tab:downstream_results_ieee}
\vspace{-0.2in}
\end{table*}

\begin{table*}[t!]
\caption{Comparison of the proposed JASPER with baseline models under the 2000-hour pre-training setup. All results are reported as classification accuracy (\%).}
\vspace{-1.8em}
\begin{center}
\resizebox{0.95\linewidth}{!}{
\begin{tabular}{|l|c|c||c|c|c|c||c|c|c||c|}
\midrule
\midrule
\multirow{2}{*}{\textbf{Model}} &
\multirow{2}{*}{\textbf{\#Params}} &
\multirow{2}{*}{\textbf{Data (hrs)}} &
\multicolumn{4}{c||}{\textbf{Audio}} &
\multicolumn{3}{c||}{\textbf{Speech}} &
\multirow{2}{*}{\textbf{Avg.}} \\
\cmidrule{4-10}
& & &
\textbf{ESC-50} &
\textbf{US8k} &
\textbf{GTZAN} &
\textbf{FMA\_SMALL} &
\textbf{IEMOCAP} &
\textbf{VoxC1} &
\textbf{SPCV2} &
\\

\midrule

HuBERT\cite{hubert}     & 95M   & 1k   & 77.45 & 77.61 & 74.00 & 42.25 & 68.30 & 86.68 & 96.10 & 74.63 \\
SSAST\cite{ssast}      & 89M   & 5.8k & 37.85 & 54.87 & 77.10 & 49.38 & 55.21 & 15.62 & 37.87 & 46.84 \\
BEATs\cite{chen2022beats}      & 91M   & 5.8k & 92.45 & 83.21 & 89.01 & \textbf{58.13} & 67.61 & 56.84 & 92.30 & 77.08 \\
EnCodecMAE\cite{pepino2025encodemae} & 86.6M & 11k & 88.25 & 85.42 & 87.10 & 56.20 & 67.80 & 71.65 & \textbf{97.30} & 79.10 \\
USAD\cite{chang2025usad}   &   94M     & 161k     & 87.45 & 84.09 & 87.05 & 54.63 & \textbf{70.61} & 69.85 & 96.65 & 78.62 \\
ATST\cite{li2023self}      & 86M    & 5.3k     & \textbf{92.90} & \textbf{87.60} & 90.01 & 55.25 & 67.79 & 68.29 & 92.53 & 79.20 \\
ULTRAS\cite{ultras}    & 89M   & 2k   & 91.15 & 86.07 & \textbf{92.59} & 55.62 & 67.78 & 73.55 & 95.10 & 80.27 \\
\midrule
JASPER     & 95M   & 2k   & 88.05 & 83.66 & 91.99 & 53.75 & 68.76 & \textbf{91.18} & 94.78 & \textbf{81.74} \\

\midrule
\midrule
\multicolumn{11}{l}{\footnotesize \textbf{US8k}: UrbanSound8K; \textbf{FMA}: Free Music Archive; \textbf{VoxC1}: VoxCeleb1; \textbf{SPCV2}: Speech Commands V2.} \\
\end{tabular}
}
\end{center}
\label{tab:baseline_comparison_2000h}
\vspace{-0.2in}
\end{table*}

\begin{table}[t!]
\caption{Comparison of phoneme recognition (PR) and automatic speech recognition (ASR) performance. In the case of HuBERT (Ours), the pre-training data is a mixture of speech and audio (2000h) similar to the JASPER, while the official implementation uses only $1000$h of Librispeech data . }
\vspace{-1.7em}
\begin{center}
\resizebox{\columnwidth}{!}{%
\begin{tabular}{|l|c|c|c|c|}
\midrule
\midrule
\multirow{2}{*}{\textbf{Model}} & \textbf{TIMIT} & \multicolumn{3}{c|}{\textbf{LibriSpeech}} \\
\cmidrule(lr){2-2} \cmidrule(lr){3-5}
 & \textbf{PER (\%)} & \textbf{PER (\%)} & \textbf{CER (\%)} & \textbf{WER (\%)} \\
\midrule
HuBERT (Official) & \textbf{11.9} & 7.1          & \textbf{2.4} & \textbf{8.5} \\
HuBERT (Ours)           & 13.0          & 7.8          & 3.7          & 12.4 \\
JASPER            & \textbf{11.9} & \textbf{6.9} & 3.3          & 10.9 \\
\midrule
\midrule
\multicolumn{5}{l}{\footnotesize \textbf{PER}: Phoneme Error Rate; \textbf{CER}: Character Error Rate; \textbf{WER}: Word Error Rate} \\
\end{tabular}
}
\end{center}
\label{tab:asr_pr_results}
\vspace{-0.2in}
\end{table}

\begin{table}[t!]
\caption{Weighted average performance across tasks under MLP and k-NN evaluation settings under XARES evaluation Benchmark (Higher values indicate better performance)}
\vspace{-1.2em}
\begin{center}
\small

\begin{tabular}{|l|c|c|}
\midrule
\midrule
\textbf{Model} & \textbf{MLP} & \textbf{k-NN} \\
\midrule

ATST-Clip\cite{li2023self}  & 0.481 & 0.433 \\
ATST-Frame\cite{li2023self} & 0.622 & 0.437 \\
BEATs\cite{chen2022beats}      & 0.451 & 0.361 \\
BYOL-A\cite{niizumi2023masked}    & 0.487 & 0.239 \\
BYOL-S\cite{turian2022hear}    & 0.391 & 0.259 \\
CED\cite{dinkel2024ced}       & 0.496 & \textbf{0.548} \\
Dasheng\cite{dinkel2024scaling}    & 0.696 & 0.499 \\
data2vec\cite{baevski2022data2vec}  & 0.565 & 0.379 \\
HuBERT\cite{hubert}   & 0.602 & 0.207 \\
MSM-MAE\cite{niizumi2023masked}    & 0.444 & 0.437 \\
w2v2\cite{wav2vec2}   & 0.384 & 0.254 \\
WavLM\cite{chen2022wavlm}      & 0.525 & 0.437 \\
Whisper\cite{radford2023robust}    & 0.646 & 0.301 \\

\midrule
JASPER      & \textbf{0.702} & 0.476 \\

\midrule
\midrule
\multicolumn{3}{l}{\footnotesize Results of standard models are tabulated from the}\\
\multicolumn{3}{l}{\footnotesize official X-ARES benchmark \cite{zhang2025x}}
\end{tabular}
\end{center}
\label{tab:mlp_knn_results}
\vspace{-0.25in}
\end{table}

\begin{table*}[t!]
\centering
\caption{Comparison of JASPER encoder with other baseline models - Track A (Higher value is better for all tasks)}
\label{tab:xares-llm-tracka-all}
\vspace{-1em}
\small % Sets a consistent, readable font size
\setlength{\tabcolsep}{3.5pt} % Adjusts horizontal padding to fit the page
\resizebox{\linewidth}{!}{
\begin{tabular}{l ll ||cccccccc|| cccc || ccc || c}
\toprule
\textbf{Model} & \textbf{Params} & \textbf{Hrs} & \multicolumn{8}{c}{\textbf{Speech Tasks}} & \multicolumn{4}{c}{\textbf{Audio Tasks}} & \multicolumn{3}{c}{\textbf{Music Tasks}} & \textbf{Avg.} \\ 
\cmidrule(lr){4-11} \cmidrule(lr){12-15} \cmidrule(lr){16-18}
& & & \textbf{SC} & \textbf{LibriC.} & \textbf{VoxL33} & \textbf{VoxC1} & \textbf{ASV15} & \textbf{FSC} & \textbf{VocalS.} & \textbf{CREMA-D} & \textbf{ESC-50} & \textbf{FSD50k} & \textbf{US8k} & \textbf{FSD18} & \textbf{GTZAN} & \textbf{NSynth} & \textbf{FMA} & \\
& & & \scriptsize (Acc) & \scriptsize (Acc) & \scriptsize (Acc) & \scriptsize (Acc) & \scriptsize (Acc) & \scriptsize (Acc) & \scriptsize (Acc) & \scriptsize (Acc) & \scriptsize (Acc) & \scriptsize (mAP) & \scriptsize (Acc) & \scriptsize (mAP) & \scriptsize (Acc) & \scriptsize (Acc) & \scriptsize (Acc) & \\
\midrule
HuBERT \cite{hubert}& 94.7M & 1k & 0.65 & 0.41 & 0.72 & 0.72 & 0.90 & \textbf{0.99} & 0.88 & 0.54 & 0.39 & 0.05 & 0.54 & 0.32 & 0.69 & 0.57 & 0.52 & 0.59 \\
WavLM \cite{chen2022wavlm} & 94.7M & 60k & 0.67 & 0.31 & 0.74 & 0.68 & 0.90 & \textbf{0.99} & 0.86 & 0.57 & 0.34 & 0.04 & 0.48 & 0.25 & 0.62 & 0.52 & 0.46 & 0.56 \\
Whisper \cite{radford2023robust}& 74.0M & 680k& \textbf{0.69} & 0.41 & \textbf{0.83} & 0.76 & 0.94 & 0.82 & 0.87 & 0.52 & 0.63 & \textbf{0.09} & 0.74 & \textbf{0.55} & 0.70 & 0.64 & \textbf{0.58} & 0.65 \\
Dasheng \cite{dinkel2024scaling}& 86.4M & 272k  & 0.65 & 0.39 & 0.31 & \textbf{0.97} & 0.94 & 0.98 & 0.85 & \textbf{0.62} & \textbf{0.75} & 0.06 & \textbf{0.83} & 0.41 & 0.32 & \textbf{0.67} & 0.43 & 0.61 \\
w2v2  \cite{wav2vec2}  & 94.4M & 1k & 0.15 & \textbf{0.47} & 0.23 & 0.60 & 0.81 & 0.33 & 0.86 & 0.42 & 0.32 & 0.04 & 0.43 & 0.25 & 0.65 & 0.48 & 0.47 & 0.43 \\
ULTRAS \cite{ultras}  & 89M & 2k & 0.52 & 0.46 & 0.28 & 0.84 & \textbf{0.95} & 0.86 & 0.91 & 0.52 & 0.60 & 0.08 & 0.79 & 0.5 & 0.78 & 0.62 & 0.56 & 0.62 \\
\midrule
JASPER & 95.0M & 2k & \textbf{0.69} & 0.45 & 0.79 & 0.74 & \textbf{0.95} & \textbf{0.99} & \textbf{0.90} & 0.55 & 0.56 & 0.07 & 0.76 & 0.43 & \textbf{0.80} & 0.63 & 0.57 & \textbf{0.66} \\
\bottomrule
\addlinespace
\multicolumn{19}{l}{\footnotesize \textbf{SC}: Speech Commands; \textbf{LibriC.}: LibriCount; \textbf{VoxL33}: VoxLingual 33; \textbf{VoxC1}: VoxCeleb1; \textbf{ASV15}: ASVspoof 2015; \textbf{FSC}: Fluent Speech Commands; \textbf{VocalS.}: VocalSound; \textbf{US8k}: UrbanSound8K;  } \\
\multicolumn{19}{l}{\footnotesize  \textbf{FMA}: Free Music Archive; \textbf{Acc}: Accuracy; \textbf{mAP}: Mean Average Precision.} \\
\end{tabular}}
\label{tab:xares-llm_tracka}
\end{table*}

\begin{table*}[htbp]
\centering
\caption{Comparison of JASPER encoder with other baseline models - Track B (Higher value is better for all tasks)}
\label{tab:xares-llm-trackb-all}
\vspace{-1em}
\resizebox{0.95\linewidth}{!}{%
\begin{tabular}{l|c||c|c|c|c|c|c||c}
\midrule
\midrule
\textbf{Model} & \textbf{Params} & \textbf{Hrs} & \makecell{\textbf{AISHELL-1} \\ \scriptsize (iWER)} &
\makecell{\textbf{Clotho} \\ \scriptsize (FENSE)} &
\makecell{\textbf{LibriSpeech} \\ \scriptsize (iWER)} &
\makecell{\textbf{MECAT} \\ \scriptsize (DATE)} &
\makecell{\textbf{Song-Describer} \\ \scriptsize (FENSE)} &
\textbf{Avg.} \\
\midrule
HuBERT   \cite{hubert} & 94.7M & 1k& 0.14 & 0.27 & 0.78 & 0.58 & 0.42 & 0.44 \\
WavLM \cite{chen2022wavlm} & 94.7M & 60k& 0.19 & 0.21 & \textbf{0.84} & 0.58 & 0.45 & 0.45 \\
Whisper   \cite{radford2023robust} & 74.0M & 680k & \textbf{0.36} & \textbf{0.36} & 0.38 & \textbf{0.62} & 0.45 & 0.40 \\
Dasheng   \cite{dinkel2024scaling} &  86.4M & 272k & 0.02 & 0.21 & 0.10 & 0.60 & 0.41 & 0.27 \\
w2v2  \cite{wav2vec2} & 94.4M & 1k & 0.0 & 0.26 & 0.0 & 0.54 & 0.43 & 0.25 \\
\midrule
JASPER & 95.0M & 2k & 0.27 & 0.35 & 0.81 & 0.61 & \textbf{0.46} & \textbf{0.50} \\
\midrule
\midrule
\multicolumn{9}{l}{\scriptsize \textbf{iWER}: $1$ - WER}; \textbf{FENSE}:Fluency-ENhanced Sentence-bert Evaluation; \textbf{DATE}: Discriminative-Enhanced Audio Text Evaluation  % Footnote added here
\end{tabular}%
}
\end{table*}

\section{Evaluation Setup}
We evaluate the proposed audio encoder through a comprehensive suite of downstream tasks and evaluation protocols. First, we assess the quality of the learned representations via a linear evaluation framework inspired by the \textsc{SUPERB} benchmark~\cite{superb}. Under this protocol, the pre-trained encoder remains frozen and serves solely as a feature extractor. Representations across all transformer layers are aggregated using a learned weighted sum, and only the aggregation weights and a lightweight, task-specific prediction head are optimized during evaluation. This approach measures representation quality independent of task-specific fine-tuning. We evaluate the model on diverse downstream tasks spanning speech, environmental sound, and music domains, covering both clip-level and frame-level prediction settings. Specifically, the model is evaluated on the ESC-50\cite{esc50}, UrbanSound8K (US8k)\cite{urbsound}, GTZAN\cite{gtzan}, FMA Small\cite{fma}, IEMOCAP\cite{iemocap}, VoxCeleb1 (VoxC1)\cite{voxceleb}, Speech Commands V2\cite{speechcommands}, CREMA-D\cite{cremad}, LibriCount\cite{libricount}, TIMIT\cite{timit} and LibriSpeech\cite{librispeech} datasets. Performance is reported using standard task-specific metrics, including classification accuracy, Phoneme Error Rate (PER), Word Error Rate (WER), and Character Error Rate (CER).

We further benchmark the effectiveness of the learned representations using two complementary evaluation settings: (i) encoder-centric evaluation using the X-ARES benchmark \cite{zhang2025x}, and (ii) audio-LLM suitability evaluation using the XARES-LLM framework \cite{interspeech2026_aecc}. The X-ARES encoder-centric evaluation employs two distinct methodologies: Multi-Layer Perceptron (MLP) linear probing and $k$-Nearest Neighbors ($k$-NN) non-parametric evaluation \cite{zhang2025x}. For the MLP evaluation, the encoder and task-specific components remain frozen, and only a single linear layer is trained to assess representation quality. Conversely, the $k$-NN evaluation is entirely non-parametric, classifying extracted embeddings based solely on their proximity within the feature space.

The audio-LLM suitability evaluation is conducted by routing the X-ARES pipeline through a Large Audio Language Model (LALM) backbone \cite{interspeech2026_aecc}. In this setting, the frozen encoder embeddings provide the acoustic representations that are used to guide an autoregressive text generator across open-ended classification and captioning tasks. The language modeling core of the XARES-LLM pipeline is driven by SmolLM2-135M \cite{allal2025smollm2}, a highly optimized, lightweight language model trained extensively on curated textual corpora. 
For the XARES-LLM  evaluation, downstream performance is mapped across two distinct evaluation pathways: 
\subsection{Track A - Classification Tasks: }
This track evaluates encoder representations on utterance-level classification tasks in domains including speech, environmental sounds, and music.
Performance is measured using accuracy (Acc) and mean average precision (mAP).
\subsection{Track B - Understanding Tasks: }
This track assesses higher-level semantic understanding using generative and sequence prediction tasks like automatic speech recognition, audio captioning, and music captioning.
Output quality is assessed using inverted Word Error Rate (iWER), Fluency-ENhanced Sentence-bert Evaluation (FENSE), and Discriminative-Enhanced Audio Text Evaluation (DATE).

\begin{table*}[t!]
\caption{Results on Track A tasks under the XARES-LLM framework for models pre-trained on $200$ hour and $2000$ hour setup. Models are evaluated using identical downstream fine-tuning and evaluation protocols. (Higher value is better for all tasks)}
\vspace{-1.7em}
\begin{center}
\resizebox{\linewidth}{!}{
\begin{tabular}{|l|c|c|c|c|c|c|c|c||c|c|c|c||c|c|c|c|}
\midrule
\midrule
\multicolumn{17}{|c|}{\textbf{200 hour setup}} \\ \midrule
\multicolumn{1}{|c|}{\multirow{2}{*}{\textbf{Model}}} & \multicolumn{8}{c||}{\textbf{Speech Tasks}} & \multicolumn{4}{c||}{\textbf{Audio Tasks}} & \multicolumn{3}{c|}{\textbf{Music Tasks}} & \multirow{2}{*}{\textbf{~Avg.~}}\\ \cmidrule{2-16}
& \makecell{\textbf{SC} \\ \scriptsize (Acc)} &
\makecell{\textbf{LibriC.} \\ \scriptsize (Acc)} &
\makecell{\textbf{VoxL33} \\ \scriptsize (Acc)} &
\makecell{\textbf{VoxC1} \\ \scriptsize (Acc)} &
\makecell{\textbf{ASV15} \\ \scriptsize (Acc)} &
\makecell{\textbf{FSC} \\ \scriptsize (Acc)} &
\makecell{\textbf{VocalS.} \\ \scriptsize (Acc)} &
\makecell{\textbf{CREMA-D} \\ \scriptsize (Acc)} &
\makecell{\textbf{ESC-50} \\ \scriptsize (Acc)} &
\makecell{\textbf{FSD50k} \\ \scriptsize (mAP)} &
\makecell{\textbf{US8k} \\ \scriptsize (Acc)} &
\makecell{\textbf{FSD18} \\ \scriptsize (mAP)} &
\makecell{\textbf{GTZAN} \\ \scriptsize (Acc)} &
\makecell{\textbf{NSynth} \\ \scriptsize (Acc)} &
\makecell{\textbf{FMA} \\ \scriptsize (Acc)} &\\
\midrule
HuBERT \cite{hubert} & \textbf{0.43} & 0.44 & 0.21 & 0.68 & 0.89 & 0.63 & 0.87 & 0.50 & 0.52 & 0.05 & 0.65 & 0.38 & \textbf{0.75} & 0.61 & 0.52 & 0.54 \\
JASPER & \textbf{0.43} & \textbf{0.45} & \textbf{0.26} & \textbf{0.85} & \textbf{0.96} & \textbf{0.72} & \textbf{0.88} & \textbf{0.52} & \textbf{0.53} & \textbf{0.06} & \textbf{0.72} & \textbf{0.42} & 0.70 & \textbf{0.64} & \textbf{0.55} & \textbf{0.58} \\
\midrule
\midrule
\multicolumn{17}{|c|}{\textbf{2000 hour setup}} \\ \midrule
HuBERT \cite{hubert} & \textbf{0.70} & 0.43 & 0.78 & 0.51 & \textbf{0.96} & \textbf{0.99} & 0.89 & 0.52 & 0.51 & \textbf{0.07} & 0.72 & 0.39 & \textbf{0.82} & \textbf{0.63} & \textbf{0.59} & 0.63 \\
JASPER & 0.69 & \textbf{0.45} & \textbf{0.79} & \textbf{0.74} & 0.95 & \textbf{0.99} & \textbf{0.90} & \textbf{0.55} & \textbf{0.56} & \textbf{0.07} & \textbf{0.76} & \textbf{0.43} & 0.80 & \textbf{0.63} & 0.57 & \textbf{0.66} \\
\midrule
\midrule
\multicolumn{17}{l}{\footnotesize
\makecell[l]{
\textbf{SC}: Speech Commands; \textbf{LibriC.}: LibriCount; \textbf{VoxL33}: VoxLingual 33; \textbf{VoxC1}: VoxCeleb1; \textbf{ASV15}: ASVspoof 2015; \textbf{FSC}: Fluent Speech Commands; \textbf{VocalS.}: VocalSound; \textbf{US8k}: UrbanSound8K;  \\ 
\textbf{FMA}: Free Music Archive; \textbf{Acc}: Accuracy; \textbf{mAP}: Mean Average Precision.
}}
\end{tabular}}
\label{tab:xares-llm-tracka-200h-2kh}
\end{center}
\vspace{-1.65em}
\end{table*}

\section{Results and Discussion}

\subsection{Linear Evaluation}
\subsubsection{Clip-level Classification}
Table~\ref{tab:downstream_results_ieee} reports the classification accuracy  of different self-supervised audio models evaluated on nine diverse downstream tasks under the two pre-training data regimes: $200$ hours and $2000$ hours. All models are pretrained on the same dataset and follow the open-source pre-training recipes provided in their respective repositories. We compare HuBERT~\cite{hubert} which uses two-stage temporal prediction-based pre-training, SSAST~\cite{ssast} trained with a masked-language modeling (MLM) objective instead of the reconstruction and location-identification losses proposed in the original framework \cite{ssast}, and JASPER. The original loss framework for SSAST has been changed for ensuring a fair comparison with the other models. Results show that the JASPER framework consistently outperforms both spectral and temporal baselines across the $200$-hour and $2000$-hour pre-training regimes on the majority of audio and speech tasks.\\
Table~\ref{tab:baseline_comparison_2000h} compares JASPER with existing state-of-the-art audio self-supervised models. Although JASPER is pretrained with only $2000$ hours of data, which is much lower than several high-compute baselines, it remains highly competitive across both audio and speech tasks and achieves the best overall average accuracy of $81.74\%$. These results highlight that the proposed joint spectro-temporal masking strategy enables JASPER to learn efficient and transferable representations with substantially less pre-training data.

\subsubsection{Frame-level Sequence Prediction}
To evaluate whether the encoder preserves fine-grained temporal information, we test frozen representations on frame-level sequence-to-sequence tasks using a lightweight linear CTC head. Table~\ref{tab:asr_pr_results} reports phoneme recognition (PR) on TIMIT and LibriSpeech under linear evaluation, and ASR on LibriSpeech-100 using a BiLSTM decoder, with CER and WER as metrics. The results show that JASPER retains detailed speech-level information despite adding spectral prediction and audio-domain pre-training objectives.

\subsection{Encoder-Centric Benchmark Evaluation (X-ARES)}
As presented in Table~\ref{tab:mlp_knn_results}, JASPER achieves best performance under the parametric MLP probing setting. In the non-parametric $k$-NN setting, JASPER retains a highly robust score, outperforming traditional speech baselines like HuBERT \cite{hubert}. The competitive performance across the parametric and non-parametric evaluations demonstrates that our joint spectro-temporal objective yields a well-structured representation space that naturally clusters by semantic similarity without relying on classifier training.

\subsection{XARES-LLM Evaluation}
Tables~\ref{tab:xares-llm-tracka-all} and \ref{tab:xares-llm-trackb-all} present the XARES-LLM results for tasks routed through the LALM backbone. In Track A, JASPER achieves the best average score of $0.66$, outperforming large-scale baselines such as Whisper\cite{radford2023robust} despite using far less pre-training data ($2$k vs. $680$k hours). While Dasheng\cite{dinkel2024scaling} and Whisper show task-specific strengths, JASPER provides a more balanced cross-domain performance profile. In Track B, JASPER again leads with an average score of $0.50$, surpassing WavLM\cite{chen2022wavlm} and HuBERT\cite{hubert}. These results suggest that JASPER's joint spectro-temporal representations provide strong and transferable features for speech and audio understanding.

\section{Ablation Studies}
\subsection{Scaling Improvements}
We study the effect of pre-training scale by training JASPER with 200 hours and 2000 hours of unlabeled speech-audio data under the same architecture and objective. Increasing the pre-training data consistently improves downstream performance across speech, environmental audio, and music tasks as shown in Table~\ref{tab:downstream_results_ieee}. This confirms that the joint spectro-temporal objective is scalable. We further analyze scaling under the XARES-LLM protocol in Table~\ref{tab:xares-llm-tracka-200h-2kh}, where increasing JASPER pre-training from 200 hours to 2000 hours leads to consistent gains when the encoder is used as the acoustic front-end of an audio-language model.

\subsection{Effect of Loss Weight $\lambda$}
To investigate the sensitivity of JASPER to the loss weight $\lambda$ in Eq.~\eqref{lambda_eqn}, we conduct an ablation study over different values of $\lambda$.
As shown in Table~\ref{tab:ablation_lambda}, the best performance is obtained at $\lambda=0.01$, where the spectral component receives a stronger weight while retaining a small contribution from the temporal objective. This trend is consistent with our central hypothesis: adding spectral prediction to a waveform encoder provides useful frequency-domain supervision, leading to stronger performance on audio tasks. At the same time, the   temporal loss preserves speech-relevant sequential information, enabling competitive performance on speech tasks.

\begin{table}[t!]
\caption{Performance of JASPER for different $\lambda$ values in the 200\,hr setup. Results are reported as classification accuracy (\%).}
\vspace{-2.2em}
\begin{center}
\renewcommand{\arraystretch}{1.25}
\setlength{\tabcolsep}{6pt}
\resizebox{\columnwidth}{!}{%
\begin{tabular}{@{}c c c c c c c@{}}
\toprule
\multirow{2}{*}{$\boldsymbol{\lambda}$} &
\multicolumn{3}{c}{\textbf{Audio Tasks}} &
\multicolumn{3}{c}{\textbf{Speech Tasks}} \\
\cmidrule(lr){2-4} \cmidrule(lr){5-7}
&
\textit{ESC-50} & \textit{US8k} & \textit{GTZAN} &
\textit{IEMOCAP} & \textit{VoxC1} & \textit{SPCV2} \\
\midrule
\textbf{0.01} & \textbf{83.55} & \textbf{78.84} & \textbf{89.69} & \textbf{65.02} & \textbf{53.93} & \textbf{92.80} \\
0.25          & 82.10          & 78.05          & 88.99          & 63.85          & 53.12          & 91.95 \\
0.50          & 80.30          & 77.83          & 88.89          & 63.46          & 51.23          & 91.23 \\
0.75          & 79.70          & 76.65          & 88.99          & 63.19          & 51.67          & 91.14 \\
\bottomrule
\end{tabular}
}
\end{center}
\vspace{-0.3em}
{\footnotesize \textbf{US8k}: UrbanSound8K; \textbf{VoxC1}: VoxCeleb1; \textbf{SPCV2}: Speech Commands V2.}
\label{tab:ablation_lambda}
\vspace{-0.6cm}
\end{table}

 \section{Conclusion}
 We introduced {JASPER}, a unified self-supervised framework for joint audio-speech representation learning. By integrating complementary time-frequency predictive objectives with temporal modeling, JASPER unifies speech and generic audio characteristics within a single raw-waveform encoder. 
 Extensive benchmarking across a diverse suite of SUPERB-style diagnostic tasks and the XARES-LLM pipeline demonstrates that the  continuous representations extracted by JASPER consistently outperform other established baselines across both frame-level and utterance-level tasks. These results highlight the benefits of unified spectro-temporal modeling for capturing the complex acoustic and semantic information required for large audio language models. 
Furthermore, all the improvements in the proposed framework are achieved with significantly smaller amounts of training data compared to several other baseline models. 

\section{AI Generated Content Disclosure}
All the scientific content, methodology, analysis, results and conclusions were developed by the authors who take full responsibility of the work. Generative AI tools were used only for minor grammatical corrections to improve the readability of the manuscript.

\bibliographystyle{IEEEtran}
\bibliography{IEEEexample} 
\end{document}